# Characterization of self-phase modulation in liquid filled hollow core photonic band gap fibers

Minh Chau Phan Huy, Alexandre Baron, Sylvie Lebrun, Robert Frey, Philippe Delaye

*Laboratoire Charles Fabry de l'Institut d'Optique, CNRS, Univ Paris-Sud, Campus Polytechnique RD128, 91127 Palaiseau cedex, France*

**Abstract :**

We experimentally and theoretically study self-phase modulation by Kerr effect in a liquid filled hollow core photonic crystal fiber. We perform a complete characterization of the linear optical properties of the hollow core photonic band gap fiber filled with deuterated acetone to determine all the characteristics of the propagation mode. The nonlinear coefficient of the fiber is determined by fitting the output spectra broadened by self-phase modulation with a new analytical expression giving the spectra of a secant hyperbolic pulse transmitted through a Kerr media. The experiment allows a precise determination of the nonlinear index change $n_2^I$ of acetone-d6 equal to $(1.15\pm0.17)\times10^{-19}$ $m^2W^{-1}$.

The advances in the manufacturing of photonic crystal fibers [1] have opened the way to the development of new devices, with optimized performances. Among the different domains that profit from the better control of holey fiber characteristics, the field of nonlinear optics has seen major advances. The small dimension reached for the silica core fibers [2] allows to obtain a very small effective mode area (of the order of a few $\mu m^2$) leading to values of the nonlinear coefficient γ up to several tens of $W^{-1}km^{-1}$ [3]. Associated to a zero dispersion wavelength displaced towards the visible spectrum, such small effective areas have lead to the development of efficient supercontinuum sources [3] that are now available commercially. Nevertheless the nonlinear performances of these fibers remain limited due to the use of silica as the nonlinear media. This problem has recently been adressed by the use of highly nonlinear glasses, such as chalcogenide glasses, for the realization of photonic crystal fibers that allow to increase the value of γ largely above thousands of $W^{-1}km^{-1}$ [4].

Another solution to overcome the limitations of silica, is to use hollow core photonic crystal fibers (HC-PCF) and fill them with gases or liquids. Compared to glass core fibers, gas or liquid filled HC-PCFs present the advantage of being very versatile. A given fiber structure can be used with several kinds of gases or liquids and give birth to a multitude of fiber devices with various properties [5]. Here again the field of nonlinear optics is particularly favored by liquid or gas-filled HC-PCFs, opening the way to new nonlinear cell development, as has already been shown with Raman converters. These hybrid cells combine the characteristics of gaseous and liquid nonlinear media (large Raman shift, small Raman linewidth, high nonlinear susceptibility, …) and the advantages of waveguided propagation (small mode diameter, large propagation length, …). Until recently research for this architecture focus on stimulated Raman scattering [6-10], and only very limited studies ever consider other nonlinear mechanisms such as Kerr nonlinearity [11-13] or Electromagnetically Induced Transparency [5, 14], with their numerous applications such as self-phase modulation, optical switching, supercontinuum generation, slow light production, and four wave mixing.

Beyond our first Raman experiments [8, 9, 15, 16], we provide a complete characterization of the Kerr effect in a liquid filled HC-PCF with a self-phase modulation experiment. This study is the basis for future experiments on the Kerr effect such as soliton production, continuum generation or parametric generation through four wave mixing. Besides signal processing applications, this fiber structure also provides a very simple and powerful characterization tool to measure nonlinear properties of liquids and gases, particularly to measure their nonlinear index change $n_2^I$. It can be used for high index liquids, using index guiding structures [12], as well as for low index liquids (or gases) using photonic band gap guiding [9], as we will show in this article. The fiber structure enables us to make use of most of the silica fiber formalism in the description of nonlinear mechanims [17], and to apply it with only minor adaptation. Propagation in the fiber can be made single mode which enables perfect definition and measurement of the mode parameters (power carried, mode diameter, …) independently from the laser characteristics. The interaction length can be large and should be adapted to the nonlinear performances of the liquid or gas to obtain sufficient nonlinear interaction for a clean measurement of the nonlinear mechanism. That precise measurement of all the parameters governing the nonlinear mechanism provides an accurate determination of the nonlinear index change of the core media.

In this paper we present a complete theoretical and experimental characterization of self-phase modulation in a photonic band gap fiber filled with a nonlinear liquid. In a first part we recall the principle of self-phase modulation by Kerr effect and present a new theoretical treatment of this well-known phenomenon, through which we obtain an analytical expression for the transmitted spectrum of a secant hyperbolic pulse through a Kerr medium. We then present a characterization of the linear optical properties of the liquid filled fiber in order to measure all the parameters that influence the nonlinear performances. Finally using the theoretical modeling and the nonlinear experimental data of the fiber, we characterize the nonlinear parameter of the fiber and deduce the Kerr coefficient of the liquid. We conclude by describing the perspective of our measurements and their generalization to other liquids.

**I. Self-phase modulation in fibers**

Self-phase modulation has been extensively studied for several decades in silica fibers [17, 18]. This penomemon manifests itself by the creation of new wavelengths in the spectrum of a pulse, owing to the variation of the nonlinear index variation which follows the temporal evolution of the pulse. The amplitude of the pulse at the output of a fiber of length L is given by [17] :

$$U(t,L) = U(t,0)\, e^{-\frac{\alpha L}{2}}\, e^{i\gamma L_{eff} |U(t,0)|^2} \qquad (1)$$

in which $L_{eff} = (1 - e^{-\alpha L})/\alpha$ is the effective length of the fiber and $\alpha$ accounts for linear losses. $\gamma = 2\pi\, n_2^I/(\lambda A_{eff})$ is the nonlinear coefficient (in $W^{-1} m^{-1}$) related to the nonlinear index change $n_2^I$ (in $m^2 W^{-1}$), $\lambda$ is the wavelength of the pulse and $A_{eff}$ the effective mode area of the fiber. $U(t,0) = \sqrt{P_0}\, f(t)$ is the amplitude of the pulse at the entrance of the fiber with $P_0$ the peak power of the pulse (in W), and f(t) the temporal shape of the pulse with a characteristic pulse duration τ. The spectral amplitude of the pulse is given by the Fourier transform of U(t,L):

$$\tilde{U}(\omega,L) = TF(U(t,L)) = \int_{-\infty}^{+\infty} U(t,L)\, e^{i\omega t}\, dt \quad (2)$$

These expressions are obtained considering that the Kerr effect is the only nonlinear effect that causes self-phase modulation. This is definitely the case in silica fibers or in liquids or gases such as those that we use, but this is not necessarily true for semiconductor materials such as silicon or GaAs where free carriers generated by nonlinear absorption of the pulse energy can cause additional phase modulation [19, 20]. The second approximation made to obtain equation (1) is that the fiber dispersion can be neglected, meaning that the dispersion length $L_D = \tau^2/|\beta_2|$, with $\beta_2$ (in $s^2 m^{-1}$) the group velocity dispersion parameter, is much larger than both the fiber length L and the nonlinear length $L_{NL} = (\gamma P_0)^{-1}$.

In the general case, the spectrum of the transmitted pulse cannot be calculated analytically and only an approached expression giving the bandwidth of the transmitted spectrum is used [17, 18] :

$$\delta\omega(t) = -\gamma L_{eff} \frac{\partial}{\partial t} |U(t,0)|^2 \quad (3)$$

In some special cases, we are able to find an analytical expression of the transmitted beam spectrum. In such cases, we stress that equation (1) can be rewritten as :

$$U(t,L) = \sqrt{P_0}\, f(t)\, e^{-\frac{\alpha L}{2}} \sum_{n=0}^{+\infty} (i\gamma L_{eff} P_0)^n \frac{|f(t)|^{2n}}{n!} \quad (4)$$

which, for the amplitude of the transmitted beam expressed in the Fourier domain, provides:

$$\tilde{U}(\omega,L) = \sqrt{P_0}\, e^{-\frac{\alpha L}{2}} \sum_{n=0}^{+\infty} \frac{(i\gamma L_{eff} P_0)^n}{n!} TF\left[f(t)|f(t)|^{2n}\right] \quad (5)$$

The evaluation of $TF\left[f(t)|f(t)|^{2n}\right]$ enables us to calculate the spectrum in the form of a finite sum (stopping the summation at a sufficient order) for different shapes of the input pulse (gaussian, exponential or secant hyperbolic). Nevertheless in some special cases we can go further. Indeed in the case of a secant hyperbolic pulse $f(t) = \mathrm{sech}(t/\tau)$, such as those delivered by a mode locked laser, the Fourier transform can be written as [21] :

$$TF\left[f(t)^{2n+1}\right](\omega) = \frac{4^n\, \pi\tau}{(2n)!}\left(\frac{1}{2} + i\frac{\tau\omega}{2}\right)_n \left(\frac{1}{2} - i\frac{\tau\omega}{2}\right)_n \mathrm{sech}\left(\frac{\pi\tau\omega}{2}\right) \quad (6)$$

using the Pochhammer symbol $(a)_n=a(a+1)\ldots(a+n-1)$ [22].

Using the fact that $(2n)!=4^n(1)_n(1/2)_n$, the amplitude in the Fourier domain writes:

$$\tilde{U}(\omega,L) = \sqrt{P_0}\, e^{-\frac{\alpha L}{2}} \pi\tau\, \mathrm{sech}\left(\frac{\pi\tau\omega}{2}\right) \sum_{n=0}^{+\infty} \frac{\left(\frac{1}{2}+i\frac{\tau\omega}{2}\right)_n \left(\frac{1}{2}-i\frac{\tau\omega}{2}\right)_n}{\left(\frac{1}{2}\right)_n (1)_n} \frac{(i\gamma L_{eff} P_0)^n}{n!} \tag{7}$$

The infinite sum corresponds to the definition of the tabulated generalized hypergeometric function $_2F_2(\{a_1,a_2\},\{b_1,b_2\},z)$ [22], that gives in the end:

$$\tilde{U}(\omega,L) = \sqrt{P_0}\, e^{-\frac{\alpha L}{2}} \pi\tau\, \mathrm{sech}\left(\frac{\pi\tau\omega}{2}\right) {}_2F_2\left(\left\{\frac{1}{2}+i\frac{\tau\omega}{2},\frac{1}{2}-i\frac{\tau\omega}{2}\right\},\left\{\frac{1}{2},1\right\},i\gamma L_{eff} P_0\right) \tag{8}$$

and for the intensity spectrum of the transmitted pulse:

$$I_\omega(\omega,L) = |\tilde{U}(\omega,L)|^2 = P_0\, e^{-\alpha L}\pi^2\tau^2\, \mathrm{sech}^2\left(\frac{\pi\tau\omega}{2}\right) {}_2F_2\left(\left\{\frac{1}{2}+i\frac{\tau\omega}{2},\frac{1}{2}-i\frac{\tau\omega}{2}\right\},\left\{\frac{1}{2},1\right\},i\gamma L_{eff} P_0\right)$$
$$\times\, {}_2F_2\left(\left\{\frac{1}{2}-i\frac{\tau\omega}{2},\frac{1}{2}+i\frac{\tau\omega}{2}\right\},\left\{\frac{1}{2},1\right\},-i\gamma L_{eff} P_0\right) \tag{9}$$

This relation shows that, as expected, the self-phase modulation spectrum is symmetrical and that its effect is independant of the sign of the nonlinear coefficient when dispersion is neglected.

The main advantage of this expression compared to the classical expression (Eq. (3)), is that it enables us to fit the whole transmitted intensity spectrum giving an exploitable value of the experimental nonlinear phase shift $\varphi_{NL} = \gamma L_{eff} P_0$ at low values, contrary to experiments using the approached expression (Eq.(3)). Indeed, Eq. (3) can only be used with spectra presenting a larger than $\pi$ shift (and sometimes even $1.5\pi$) [18]. This is because, for low nonlinear phase shifts, experimental evaluation of the spectral broadening is problematic [23].

The intensity spectra (Fig.1) obtained from Eq. (9) present the classical shapes associated with self-phase modulation [17, 18]. These spectra also show a perfect match with the spectra that we obtained through a numerical nonlinear beam propagation simulation using the Split Step Fourier Method [17].

As a result, we have at our disposal a very simple analytical expression to calculate the spectral broadening of a secant hyperbolic pulse induced by self-phase modulation by the Kerr effect. Even though this is only valid for this particular temporal shape of the pulse, this model is quite general nonetheless, because most pulsed mode locked lasers deliver such kind of pulse shapes. Moreover most trends that can be extracted with this model are exploitable for other temporal pulse shapes with only small adaptations. We propose in the following to study this process in a liquid filled photonic crystal fiber.

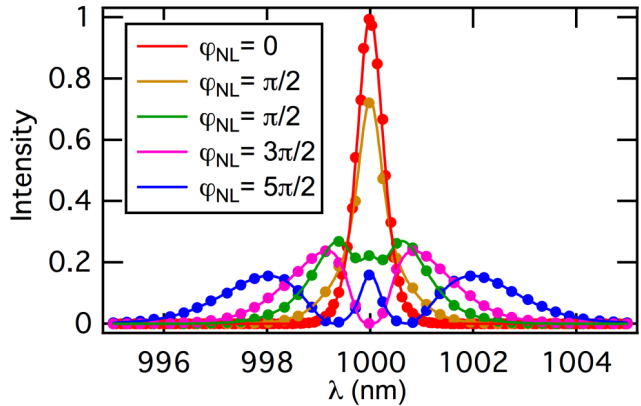

Figure 1 : Calculated intensity spectrum of a secant hyperbolic pulse through a 1m nonlinear fiber, for different values of the nonlinear phase shift (absorption is neglected for simplicity). The solid lines correspond to the analytical model (Eq. (9)) and the markers to the numerical simulation using the split step Fourier method.

**II. Linear propagation properties of the liquid filled hollow core fiber**

The spectral broadening caused by self-phase modulation depends on many fiber parameters (absorption, dispersion, effective mode area and nonlinear index change) and on the experimental set-up (work wavelength, pulse duration and incident power). Most of these parameters can be obtained experimentally by independent measurements, which we will describe. The only exception is the nonlinear index change $n_2^l$ that will be determined by careful analysis of the self-phase modulated signal.

II.1. Experimental set-up

The experimental set-up is shown on Fig. 2. The beam of a tunable mode-locked picosecond Ti:Sapphire laser is sent into a liquid filled HC-PCF where it experiences self-phase modulation by Kerr effect. The output beam (signal beam) is imaged on a CCD camera, using a microscope objective and an achromatic lens, to analyze the near field spatial structure of the mode propagating in the fiber. The same beam is also sent on a Si detector or in a fiber connected to an optical spectrum analyzer. Part of the incident beam is extracted via a beam splitter cube to form a reference beam. The reference beam is recombined with the signal beam using a second beam splitter cube. This reference beam enables us to measure the incident intensity spectrum, as well as the dispersion of the fiber using the interferometric time of flight technique [24] (connected to the detector providing a measurement of the constrast of the interference signal between the signal and the reference beams).

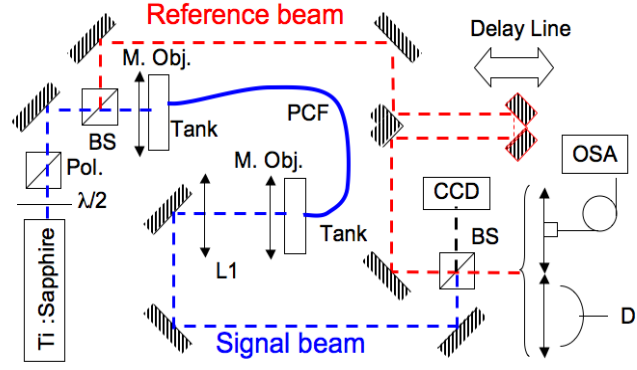

Figure 2 : Schematic experimental set-up. BS: beam splitter cube, λ/2: Half wave plate, Pol.: Polarizer, M.Obj.: x6.3 microscope objective, D: Detector, OSA: Optical spectrum analyzer, PCF: Liquid filled hollow core photonic crystal fiber.

The fiber is fixed in tanks closed by BK7 windows and all the fiber holes are totally filled using a process described in [15]. The average power of the input and output beams are measured before and after the microscope objectives respectively, using a power meter. They are corrected from the transmission of the microscope objective and of the BK7 windows of the tank to measure the power injected inside the fiber and collected at the exit facet of the fiber.

### II.2. The picosecond Ti:Sapphire laser

The laser source is a mode-locked picosecond Ti:Sapphire laser (Tsunami from Spectra-Physics) tunable between 700nm and 980nm, which delivers pulses close to 1 picosecond in duration at a repetition rate $f_R$=80MHz. The output of the laser is continuously monitored with an autocorrelator that measures $\tau_{AC}$, the full width at half maximum of the autocorrelation function, with a precision on the order of ±1 to ±2%. For a secant hyperbolic pulse ($f(t) = \mathrm{sech}(t/\tau)$) delivered by such a mode locked laser, the pulse duration is given by $\tau = \tau_{AC}/2.720$ [24]. Using this temporal shape, it is easy to link the peak power $P_0$ of the pulse to the pulse energy $E_p$ and to the average power $\overline{P}$ of the beam.

$$P_0 = \frac{E_p}{2\tau} = \frac{\overline{P}}{2\tau f_R} \qquad (10)$$

With these parameters we can also calculate the spectrum of the incident beam, given by :

$$I_\omega(\omega, L) = P_0\, e^{-\alpha L} \pi^2 \tau^2 \,\mathrm{sech}^2\!\left(\frac{\pi \tau \omega}{2}\right) \qquad (11)$$

and compare it with the measured spectrum (Fig. 3).

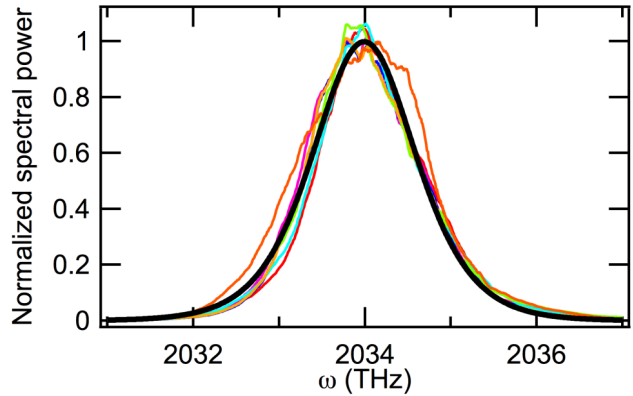

Figure 3 : Comparison between a batch of experimental spectra obtained with identical operating conditions for the laser (λ=926.8nm) and the theoretical spectrum calculated using Eq. (11) and the value of τ=0.794ps deduced from the autocorrelation measurement.

We can see on Fig. 3 the very good agreement between the theoretical spectrum and the experimental ones, despite the slight dispersion in the measured data that were obtained with identical operating conditions for the laser but at different measurement times. From the measured spectra of the incident beam, we confirm that the pulses delivered by the Ti:Sapphire laser are Fourier transformed with secant hyperbolic temporal shapes. In the following, the laser is aligned in order to deliver these secant hyperbolic Fourier transformed pulses with durations measured via the autocorrelator with a precision on the order of ±1-2% depending on the measurement.

### II.3. The liquid filled hollow core fiber

The choice of the liquid filled hollow core fiber in the present experiment is based on several constraints, such as the operating wavelength of the laser source, the nonlinear efficiency of the liquid and its ease of use, or the availability of both fiber and liquid. We choose to operate with a commercial hollow core photonic band gap fiber [5] (HC-1550-02 from Crystal Fiber) filled with deuterated acetone (acetone-d6). When the fiber is filled with a liquid with the index of refraction of acetone-d6 (around 1.35) [25, 26], the transmission band of the fiber around 1550nm is shifted towards 800nm [27]. Experimentally, the entire spectral range of the Ti:Sapphire laser is covered, as shown on Fig. 4.

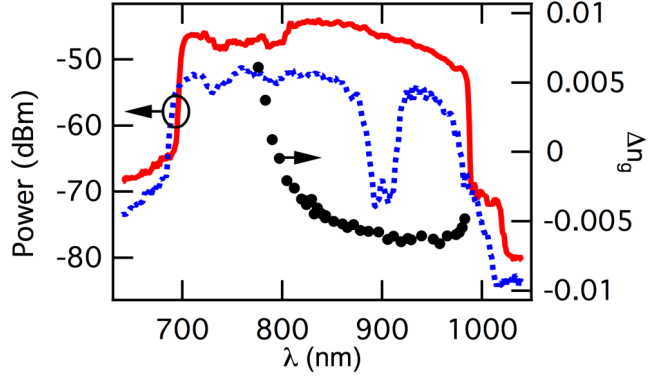

Figure 4 : Power of a supercontinuum source transmitted through the HC-1550-02 hollow core fiber filled with acetone (blue dotted curve) and acetone-d6 (red line). The additional losses due to overtones of IR absorption bands of acetone are eliminated by deuteration, giving a full transmission band for the acetone-d6 filled fiber. The shift of the lower edge of the transmission band of the fiber filled with acetone-d6 is due to the slightly lower index of acetone-d6 compared to non-deuterated acetone [25]. The black dots show the measured variation of the group index of the fiber, from which the dispersion is deduced.

The absolute transmission of the fiber is more complex to evaluate. We typically obtain a transmission greater than 50% without too much effort, without being able to completely decouple the coupling and the mode propagation losses. The best value obtained for the transmission is 87% i.e. a 0.6dB loss for the 80cm long fiber. That would correspond to an upper limit for the propagation losses equal to 0.75dB.m$^{-1}$ (i.e. $\alpha < 0.17$m$^{-1}$) considering a perfect coupling efficiency. As a result, we can reasonably consider that propagation losses are negligible on a distance of 1m of fiber, which we use in all our experiments.

The fact that propagation losses can be neglected also means that the value of the collected output power of the fiber provides a good measurement of the power propagating inside the fiber and can therefore, be used as a measurement of the incident average power $\overline{P}$ and the pulse energy $E_p$ using Eq. (10). The maximum measured transmitted average power is on the order of 40mW with a precision close to ±2% typically. This gives a pulse energy that reaches 600pJ per pulse at maximum power with the same typical error. The influence of the propagation losses on the value of the nonlinear coefficient measured in the nonlinear experiment is discussed in Appendix A.

Using the experimental set-up, we also measure the variation of the propagation time $\Delta t$ (related to the variation of position of the delay line) as a function of wavelength in the liquid filled hollow core fiber. From this measurement, we deduce the variation of group index $\Delta n_g = c\,\Delta t/L$ as a function of wavelength (Fig. 4). The fit of the $\Delta n_g$ experimental data as a function of $\omega$ by a polynomial curve provides the group velocity dispersion of the fiber $\beta_2 = (1/c)\partial \Delta n_g / \partial \omega$. We obtain $\beta_2$(in s$^2$m$^{-1}$) = $-1.304576733 \times 10^{-20} + 5.70128632 \times 10^{-17}\,\lambda^{-1}$ -

9.964374186x10$^{-14}$ λ$^{-2}$ + 8.70611331x10$^{-11}$ λ$^{-3}$ - 3.802860616x10$^{-8}$ λ$^{-4}$ + 6.6437917x10$^{-6}$ λ$^{-5}$ with λ in nm. From these data we show that the dispersion length is greater than 10m (i.e. $L_D$>>L) on the whole spectral range extending from 820nm to 970nm, with a zero dispersion wavelength located around 946nm. Therefore, we are in the conditions in which the theoretical model applies.

Inspite of the fact that propagation in the liquid filled hollow core fiber is not strictly single mode [9], the gaussian fundamental mode of the fiber can be selectively excited without much difficulty. By imaging the fiber output plane on a CCD camera (Fig. 2), we are able to determine the mode intensity radius w at 1/e$^2$ and $A_{eff}$=πw$^2$ the effective area [17] of the gaussian transmitted mode (Fig. 5). We measure at λ=926nm, w=4.4±0.2μm and $A_{eff}$=61±5μm$^2$ and at λ=951nm, w=4.2±0.2μm and $A_{eff}$=55±5μm$^2$. The mode energy is then almost completely included in the 10.9μm core diameter of the fiber.

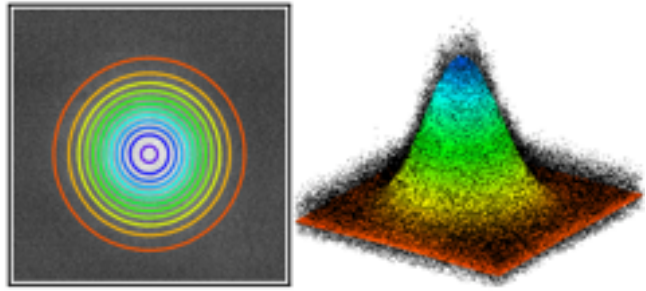

Figure 5 : Image on the CCD camera of the fundamental mode exiting the fiber (black and white) at λ=951nm, with a 2D gaussian curve fit (color).

**III. Self-phase modulation in the liquid filled hollow core photonic crystal fiber**

Once the linear characteristics of the fiber are determined, we measure and analyze the spectrum of the beam transmitted by the fiber when the pump power is increased at different wavelengths. Fig. 6 shows typical transmitted spectra at the highest power demonstrating a clear symmetrical broadening attributed to self-phase modulation by Kerr effect with a nonlinear phase shift reaching 1.5π (see Fig. 1) [17, 18]. The broadened spectra are compared with the spectra calculated from Eq. (9) derived from the analytical model, and the numerical Split Step Fourier model taking into account the dispersion of the fiber, measured previously. The agreement between the different spectra is excellent.

The influence of dispersion is not visible in the curve presented on the linear scale. If we use a Log scale (Fig. 6D), we observe a variation in the transmitted spectra due to dispersion. By comparing the different theoretical curves, we see additional components appearing on the edges of the broadened spectra at normalized intensity levels below 10$^{-3}$.

Nevertheless experimentally these effects are hardly visible as they are masked by small imperfections of the incident beam spectra and fluctuations of the transmitted spectra which are on the same order of magnitude as the expected spectral variation. We can then conclude that no clear experimental evidence of the influence of dispersion can be seen in our measurements, validating the approximation of negligible dispersion that was made in order to establish our analytical model. We can also see a small assymetry in the amplitude of the broadened peaks for some of the curves. This behaviour is not clearly reproducible and we attribute it to fluctuations in the detected signal (see Fig.3 for another example) rather than to the influence of a temporal assymetry of the incident pulse as discussed in ref. [18].

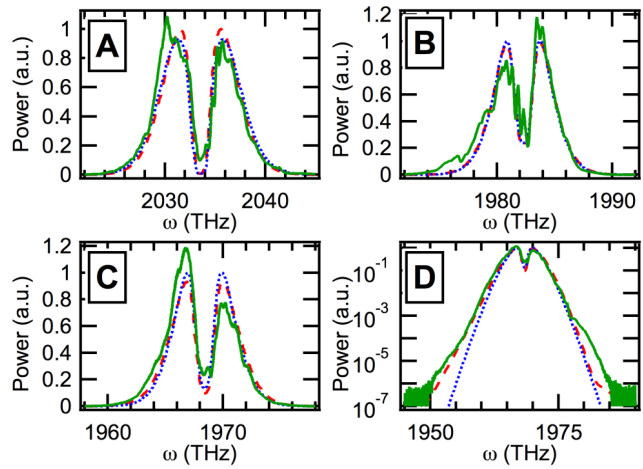

Figure 6 : Measured transmitted spectrum at different wavelengths at maximum pump power (green line): A: 926.8nm (τ=0.794ps), B: 950.8nm (τ=0.934ps), C: 957.7nm (τ=0.934ps), D corresponds to C presented in Log scale. The spectra are compared with the spectra calculated using expression (9) (blue dotted curves) and spectra given by the numerical Split Step Fourier Model taking into account the measured dispersion of the fiber (red dashed line). The pulse durations are deduced from the autocorrelation measurement.

The experimental transmitted spectra at different pulse energies are then fitted with the theoretical model to determine the nonlinear parameters of the transmitted spectrum (Fig. 7). For the fit we use the theoretical expression given by Eq. (9). As the pulse duration τ is known through the autocorrelation measurement, we only need two parameters to fit the experimental data: the nonlinear phase shift $\left(\gamma L_{eff} E_p / (2\tau)\right)$ and the amplitude of the signal which combines the normalization factor $\left(E_p\, e^{-\alpha L} \pi^2 \tau / 2\right)$ and the coupling coefficient of the transmitted signal to the optical spectrum analyzer (we use Eq. (10) to take into account the dependence of $P_0$ with the pulse duration). In practice, we note that the precision of the fitting parameters is governed by the precision on the value of the pulse duration. By varying the pulse duration within the error bars (typically 1% to 2%) and performing several fits of the experimental curves, we are able to estimate the error bars on the fitting parameters.

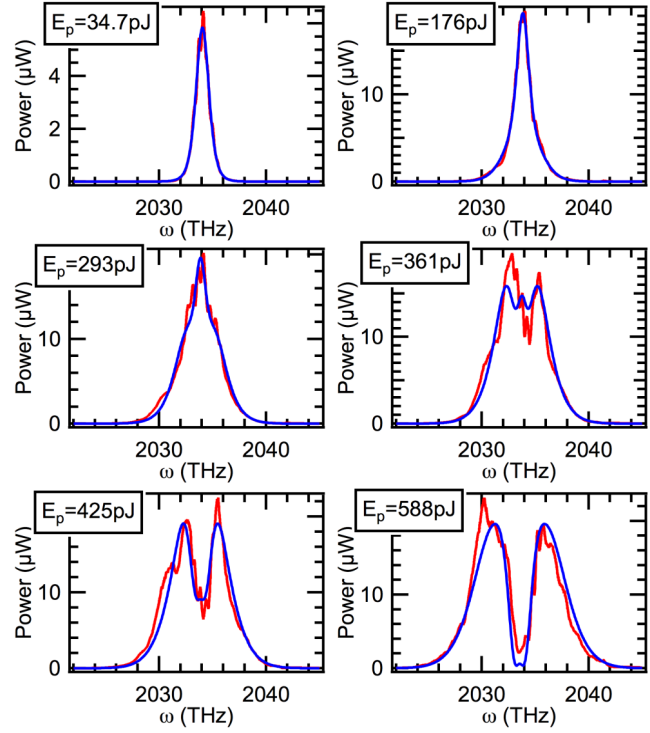

Figure 7 : Spectra of the pulses (red curve) transmitted through the liquid filled hollow core fiber and the fit (blue curve) with Eq. (9) at different incident pulse energies in the fiber at a wavelength of 926.8nm (the incident spectra are the ones shown in Fig. 3).

The quality of the fits is very good as shown on Figure 7 where we show the experimental spectra and their fits for λ=926.8nm. From these fits we determine the value of the nonlinear parameter at different energies and plot them as a function of $E_p$ (Fig. 8). The points show the expected linear dependence whose slope gives the coefficient $\gamma L_{eff} = \gamma L$. The alignment of the experimentally determined points on the fitting line is very good except for the two points at small energy at λ=957.5nm. For the point at lowest energy, the nonlinear parameter determined by the fit was zero, giving an indication of the minimal broadening necessary to realize a proper fit of the transmitted spectrum. Considering that the highest achieved nonlinear phase shift is around 1.5π, a nonlinear phase shift of the order of 0.1π to 0.2π should be sufficient for the characterization of the nonlinear performances of a sample. This is one order of magnitude lower than the classical treatment of self-phase modulation spectra based on a measurement of the frequency broadening [18].

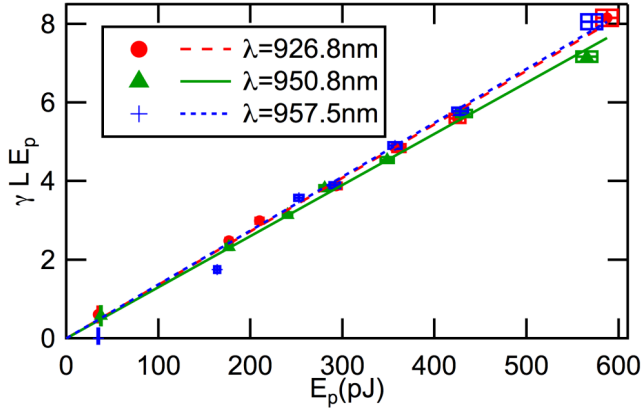

Figure 8 : Plot of the nonlinear coefficient determined by fitting the transmitted spectrum as a function of the energy of the pulse with measurements at three different wavelengths.

Using the length L=1m of the fiber, we deduce from these measurements the value of the nonlinear coefficient γ with a precision on the order of 3% to 6% depending on the quality of the data series. The results are summarized in Table I. The value of γ is on the order of 13.4 $W^{-1}km^{-1}$ for all wavelengths. Using the definition of γ we can also deduce from these measurements, the nonlinear index change $n_2^I$ of the fiber (which in our case will be the nonlinear coefficient of acetone-d6 as most of the mode energy propagates in the liquid core of the fiber) using the measured value of the effective mode area (Part II.3).

| λ (nm) | $A_{eff}$ ($\mu m^2$) | γ ($W^{-1}km^{-1}$) | $n_2^I$ ($10^{-19}$ $m^2W^{-1}$) |
|---|---|---|---|
| 926.8 | 61±5 | 13.61±0.40 | 1.22±0.17 |
| 950.8 | 55±5 | 13.01±0.34 | 1.08±0.15 |
| 957.5 | 55±5 | 13.71±0.87 | 1.14±0.16 |

Table I : Nonlinear coefficient and nonlinear index change calculated from our measurements at different wavelengths in the liquid filled hollow core fiber.

The average value of $n_2^I$ is equal to (1.15±0.17)x$10^{-19}$ $m^2W^{-1}$, which is 4 times higher than the $n_2^I$ of silica [17]. Across the wavelength range of the experiment $n_2^I$ presents no dispersion within the precision of the measurements. The precision of the measurement is mainly governed by the precision in the determination of the effective mode area (error on the order of ±8%), that is relatively high in this first experiment and can probably be improved for future measurements. No value of $n_2^I$ for acetone-d6 can be found in the literature. We can only compare our results with the $n_2^I$ value of acetone supposing that deuteration has only little influence on the Kerr effect. Our value is completely coherent with the previously determined values of $n_2^l$ for acetone considering their large dispersion, because they range from $2.4 \times 10^{-19}$ m$^2$W$^{-1}$ [28], to $0.59 \times 10^{-19}$ m$^2$W$^{-1}$ [29], with an intermediate value of $1.33 \times 10^{-19}$ m$^2$W$^{-1}$ [30].

The value of the nonlinear coefficient γ of the liquid filled fiber obtained in this first experiment is promising. Indeed, despite the fact that the liquid used here is not known for its high nonlinear susceptibility (in this proof of principle experiment, we choose it essentially for practical reasons), γ is close to the values obtained in highly nonlinear photonic crystal fibers. This is all the more encouraging, considering that this value is obtained with a fiber that has an effective area close to that of classical silica fibers, opening the way to the realization of easy-to-use devices, easy to couple to other silica fiber devices such as couplers or fiber Bragg gratings. In order to access higher nonlinearities, we can change the liquid to beneficiate from a higher susceptibility, but we can also optimize the fibers to reduce the effective mode area, which would lead to higher values of the nonlinear coefficient because of the high nonlinear susceptibility of liquids and the small dimension of the fiber core.

**IV. Conclusion**

Self-phase modulation by Kerr effect in a liquid filled hollow core photonic crystal fiber opens the way to a new and easy-to-use architecture for the measurement of the nonlinear refractive index change of liquids and gases. Compared to other techniques such as Z-scan measurements, the beam parameters are no longer controlled by the laser quality but by the modal properties of the fiber. This means, as we show in this paper, that all the parameters required for the determination of the nonlinear parameter can be measured with good precision and few assumptions. Coupled to a new analytical expression that can be used to fit the spectrum of the beam transmitted through the fiber, the experimental set-up enables us to determine the nonlinear index change of deuterated acetone with a good precision. With the preliminary set-up used in this first experiment, we provide a measured value with a precision on the order of ±15%, that is essentially limited by the quality of the measurement of the effective mode area. Further improvements of the set-up should allow us to decrease this uncertainty below ±10%, a value that should be sufficient to access the spectral variation of the nonlinear index change.

The experiment can be easily implemented with other liquids with relatively few changes in the set-up. For example, with the same fiber we used, a change of the refractive index of the liquid of ±0.03 induces a shift in the transmission band of ±100nm [27]. That band still covers at least partially the wavelength range of the Ti:Sapphire laser. The experimental set-up can then be used with a large variety of liquids, from water (with n=1.33) to propanol (with n=1.38), and intermediate liquids with higher values of $n_2^I$ such as formic acid (n=1.37) [31]. Other fibers, with the same shape or with other shapes such as Kagome or derived fibers [10, 32], as well as other types of hollow core photonic crystal fibers [12] could be used with other liquids with higher index.

Beyond these self-phase modulation experiments, knowing the nonlinear properties of liquid and gas filled hollow core photonic crystal fibers, opens the way to the demonstration and characterization of other nonlinear mechanisms, such as soliton propagation or parametric generation through four wave mixing, which are well known in silica and glass core fibers, but are still poorly studied in liquid core fibers. The large variety of liquids and gases with their specific linear and nonlinear properties opens the way to the realization of a large variety of new optimized nonlinear functions.

**Acknowledgment:**
Minh Chau Phan Huy acknowledges the region Ile-de-France for its financial support.

**Appendix A**

All the interpretations of the nonlinear experimental results have been made by neglecting the effect of propagation losses in the fiber $\alpha$. Taking into account this absorption is relatively easy, if we remember that the self-phase modulation spectral broadening is governed by a single parameter: the nonlinear phase shift $\varphi_{NL} = \gamma L_{eff} P_0$. This parameter is determined independently from the assumptions made on the value of the propagation losses. As we measure the intensity at the output of the fiber, when taking the propagation losses into account, we can write $P_0 = e^{\alpha L} P_{out}$, and the nonlinear phase shift rewrites as:

$$\varphi_{NL} = \gamma_\alpha \left( \frac{e^{\alpha L} - 1}{\alpha L} \right) L P_{out} \tag{A.1}$$

where $\gamma_\alpha$ is the nonlinear coefficient including propagation losses. It is related to the coefficient determined by neglecting the propagation losses $\gamma_0$ related to $\varphi_{NL} = \gamma_0 L P_{out}$, by:

$$\gamma_\alpha = \gamma_0 \frac{\alpha L}{e^{\alpha L} - 1} \tag{A.2}$$

For the fiber studied here, the propagation losses are at worst equal to $\alpha = 0.17 \text{m}^{-1}$. This means that $0.92\gamma_0 < \gamma_\alpha < \gamma_0$, indicating that in the worst case scenario the nonlinear coefficient is overestimated by less that 10%.

**<u>References</u>**